\documentclass[aps, prb, amssymb, amsmath, superscriptaddress, 
twocolumn] {revtex4-2}

\usepackage{chngcntr}

\usepackage{graphicx}
\usepackage{color}
\usepackage[dvipsnames]{xcolor}
\usepackage{amsmath}
\usepackage{enumitem}
\usepackage{amssymb}
\usepackage[hidelinks]{hyperref}
\usepackage{cancel}
\usepackage{ulem}%%%%%%%
\usepackage{multirow}
\newcommand{\be}{\begin{equation}}
	\newcommand{\ee}{\end{equation}}

\newcommand{\bea}{\begin{eqnarray}}
	\newcommand{\eea}{\end{eqnarray}}

\makeatother

\begin{document}
	\title{Effective model for Pb$_{9}$Cu(PO$_4$)$_6$O}
	
\author{Patrick A. Lee}
\affiliation{
Department of Physics, Massachusetts Institute of Technology, Cambridge, MA, USA
}
\author{ Zhehao Dai}
\affiliation{
Department of Physics, University of California, Berkeley, CA, USA
}

\begin{abstract}
The copper substituted Pb-apatite has attracted a great deal of attention recently, due to the claim of the observation of room temperature superconductivity. Based on LDA calculations in the literature, we propose an effective model that describes the low energy physics. It consists of stacks of buckled honeycomb lattices, with Cu and O occupying the A and B sites respectively. In addition to the narrow Cu bands that have been emphasized, we call attention to the relatively small energy separation between the Cu and O orbitals. Thus despite the small hoping energies, the model may be in an interesting regime near the metal insulator transition driven by the charge transfer mechanism. Relationships with cuprates and the organic superconductors are discussed.

\end{abstract}
%\date{\today}

\maketitle

The recent papers by S. B Lee et al \cite{lee2023firs,lee2023superconductor} reporting the observation of room temperature superconductivity (SC) in the copper substituted Pb apatite
Pb$_{10-x}$Cu$_x$(PO$_4$)$_6$O (called LK-99) has generated great excitement as well as quite a bit of skepticism. The basic electronic structures for the composition $x=1$ have been elucidated by several LDA band structure calculations which are in broad agreement. \cite{griffin2023origin,si2023electronic,kurleto2023pb,cabezas2023theoretical,lai2023first} A key finding is that the Fermi surface is dominated by two Cu d orbitals (mainly  $d_{xz}$ $d_{yz}$ ) which form two exceptionally narrow bands with total bandwidth of order 0.15 $eV$. These two bands are filled by 3 electrons, forming a metal within band theory. From the structure, it is clear that the Cu ions are far apart from each other, with a spacing of about 10 angstroms. Normally a narrow band is a signature that the electronic orbitals have a small overlap. With an odd number of electrons per unit cell, the ground state is expected to be deep in the Mott insulator limit. Thus even a metallic ground state is surprising, let alone superconductivity. This problem was emphasized in ref.\cite{si2023electronic}. One could consider deviation from $x=1$, leading to the problem of a doped Mott insulator. However, for a small $t/U$ ratio, where $t$ and $U$ are the effective hopping integral and Hubbard onsite repulsion respectively, the exchange constant $J\approx 4t^2/U$ is expected to be very small. This is unlike the case of the cuprates, where intermediate coupling leads to a large $J$ which is understood to be the driving force behind interesting correlation physics and superconductivity.\cite{lee2006doping} At first sight it seems that this particular system is unfavorable to superconductivity. The purpose of the paper is to point out that this is not necessarily so. Based on the band structure calculations we propose an effective model which could potentially harbor unconventional superconductivity, even though the $T_c$ is unlikely to be very high due to the small energy scale of the parameters. Nevertheless, we believe the effective model is a worthy one to study, regardless of whether the claim of room temperature SC will gain wide acceptance.

\begin{figure}
	\centering
	\includegraphics[width=0.5\textwidth]{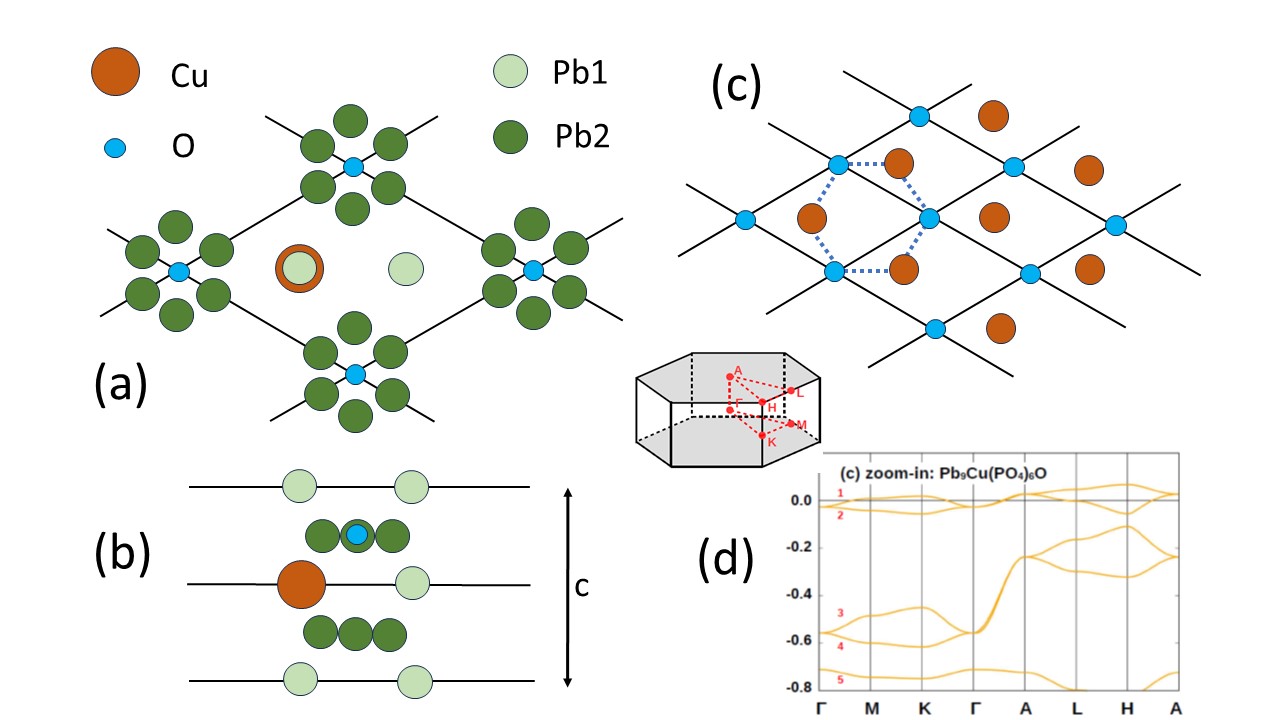}
	\caption{ 
(a) Top view and (b) side view of the structure of Pb$_{9}$Cu(PO$_4$)$_6$O, also known as LK-99, leaving out the  PO$_4$ (adapted from \cite{cabezas2023theoretical}). Six Pb2 ions form two triangles rotated by 60 degrees and separated by $c/2$ within the unit cell. Oxygen occupies the center of the top triangle as shown in (b), but an alternative site is in the center of the lower triangle. There are 4 Pb1 sites in the unit cell, one of which has been substituted by Cu.  (c) The Cu and O ions form a  stack of buckled honeycomb lattice. (d) Band structure of  Pb$_{9}$Cu(PO$_4$)$_6$O showing the mainly Cu bands (1,2) near the Fermi energy and the mainly oxygen bands (3,4) below. (reproduced from \cite{si2023electronic}) These bands are quite isolated from the remaining filled bands and represent the low energy hole excitations. }
	\label{fig1}
	\vspace{-2mm}
\end{figure}
Let us begin with a summary of the structure which has a hexagonal unit cell. We will set aside the phosphate PO$_4$ and focus on the Pb and Cu sites and the single additional oxygen in the unit cell. There are two kinds of Pb sites, called Pb1 and Pb2. Viewed from the top, 6 Pb2 forms a hexagonal structure comprising two triangles in different layers rotated by 60 degrees. (We follow the notation of ref. \cite{griffin2023origin,si2023electronic}.) These form a channel along the $c$ axis which accommodates the oxygen ion. From the side view in Fig.~\ref{fig1}b, we see that there are two inequivalent sites for the oxygen which sits in the center of either one of the two Pb2 triangles. There are 4 Pb1 sites per unit cell, one of which is substituted by Cu. We assume the Cu occupies the site shown in Fig.~\ref{fig1}a-b, as assumed by most LDA calculations.

The Cu ions are far apart, so the direct overlap is negligible, However,  they have reasonable overlap with Pb orbitals and with numerous oxygen orbitals (the shortest distance is 2.06 Angstrom.) and the wavefunctions have significant hybridization with the oxygen $p$ orbitals.\cite{si2023electronic,lai2023first,kurleto2023pb} These overlaps are responsible for the small but finite bandwidth. The band dispersion is a bit larger in plane than along the $c$ axis, so the system is not at all one-dimensional. In addition, just below the flat bands lie another two bands (labeled band 3 and 4 in ref.\cite{si2023electronic} and reproduced in Fig. \ref{fig1}d) which are mainly centered on the oxygen ion. A most remarkable feature of the electronic structure is that the middle of this band is only  0.4 $eV$ below the Cu orbitals. The bands are formed out of 2 oxygen p orbitals and this suggests that the energy separation $D$ between the oxygen and Cu orbital is only about 0.4 $eV$. For comparison, in the cuprate, this energy is about 2 $eV$. The parent compound of the cuprate is a charge transfer insulator \cite{zaanen1985band} where the energy $D$ plays the role of the Hubbard U in an effective Hubbard model. The much smaller value of $D$ in LK-99 suggests that despite the small hopping matrix element implied by the almost flat band, this system may also be in the intermediate strong correlation regime where unconventional superconductors may reside.

The above considerations lead us to propose the following effective model for LK-99 which we believe captures the low-energy physics.
As shown in Fig. \ref{fig1}c  we isolate the Cu and the oxygen ion and consider a stack of buckled honeycomb structures where Cu and O occupy the A and B sites. (The B site lies in a plane $c/4$ above the A plane occupied by Cu. Thus inversion symmetry is broken. There is another version of the structure where it lies a distance $c/4$ below.)    There are two degenerate d orbitals ( $d_1$ and $d_2$) on Cu and two degenerate p orbitals ( $p_1$ and $p_2$) on O. In the hole picture, there is a single hole in the d band and the p hole orbitals lie at an energy $D$ above the d orbitals. We keep the nearest neighbor hopping between the d and p orbitals. These layers are stacked directly on top of each other. For reasons explained below, the important hopping between the layers is the NN hopping between the oxygen p orbitals. The Hamiltonian in the hole picture is the sum of an onsite term and a hopping term. $H=H_0 + H_h$
\begin{widetext}
\be \label{H0}
H_0 =  \sum_{k,\sigma,\alpha} (\sum_{i \in A} \epsilon_d d_{\alpha,\sigma}^{\dagger}(i,k)d_{\alpha,\sigma}(i,k) + \sum_{j \in B} \epsilon_p p_{\alpha,\sigma}^{\dagger}(j,k)p_{\alpha,\sigma}(j,k))
\ee
\be \label{Hh}
H_h =  \sum_{k,\sigma} (\sum_{<ij>, \alpha,\beta} (t_{ij,\alpha \beta}d_{\alpha,\sigma}^{\dagger}(i,k) p_{\beta,\sigma}(j,k) + c.c.) + \sum_{j ,\alpha} (t_c p_{\alpha,\sigma}^{\dagger}(j,k)p_{\alpha,\sigma}(j,k+1) +c.c.)
\ee
\end{widetext}
Importantly, this Hamiltonian is supplemented by the requirement that no double occupation is allowed on the Cu sites, i.e. we set the Hubbard $U$ on the Cu sites to infinity. There is also a Hubbard repulsion $U_O$ on the oxygen site which will further impede metallicity and reduce the exchange interaction. Here we will assume it is small. In these equations,
$k$ labels the layers, and $i$ and $j$ labels the A and B sites in the honeycomb lattice. $\alpha=1,2$ and $\beta=1,2$ are orbital labels, $\sigma$ is the spin label. $d_{\alpha,\sigma}^{\dagger}(i,k)$ creates a hole in the $\alpha $ orbital on site $(i,k)$ and similarly for the p orbital. $\epsilon_p=\epsilon_d + D$. We assume that different components of $t_{ij,\alpha\beta}$ have the same order of magnitude which we denote by a single parameter $t$. 
 %{\red I don't think it's true. First $t_{\alpha,\beta}$ depends on the direction of the i-j bond, for different i and j t are related by symmetry. But for a single bond, a generic 2 by 2 t is allowed by C3 rotation, which is the only symmetry. We may say different components of $t_{\alpha,\beta}$ have the same order of magnitude.}
 This model is characterized by 3 parameters, $t,t_c$, and $D$. We will next estimate their values.

 The  Cu bands have a total bandwidth of about 130-160 meV, depending on the different versions of LDA calculations.\cite{griffin2023origin,si2023electronic} Since Cu sits on a triangular lattice, the bandwidth is 9 times the effective hopping $t_{LDA}$. Since the charge content of this band is mainly on Cu and the oxygen ion, we assume that $t_{LDA}$ comes from hopping on and off the oxygen ion and is given by 
$t_{LDA}=t^2/D$. Using D=400 meV, we estimate the typical hopping 
integral $t_{\alpha \beta}$  between Cu and O to be about 80 meV with considerable uncertainty due to the crudeness of our estimate. The hopping between Cu on different layers involves a longer path via the oxygen ion and is expected to be small.

The oxygen bands show a small dispersion in the plane and a rather large dispersion along c, from $\Gamma$ to $A $ the in-plane hop via Cu involves an intermediate state with double hole occupation on Cu and is strongly suppressed by the onsite $U$ on Cu. Thus we believe the in-plane dispersion found by LDA is an overestimate, and we can ignore the in-plane hopping between the oxygen ions. The dispersion from $
\Gamma$ to $A$ is about 400 meV and gives twice the hopping integral $t_c$. Therefore we estimate $t_c$ to be about 200 meV. We note that this band is already present in the absence of Cu substitution with a similar $
\Gamma$ to $A$ dispersion.\cite{lai2023first} Therefore this hopping integral most likely does not involve an intermediate step through the Cu site and the LDA estimate should be reliable. We take the midpoint of the dispersion to be the location of $\epsilon_p$. By assuming $\epsilon_d$ to be near the Fermi energy, we come up with the estimate $D=400 meV$ mentioned earlier.

It is worth noting that there are two potential oxygen sites located at the center of the two Pb2 triangles, only one of which is occupied. One may expect substantial disorder in the occupation of these sites. However, as far as the in-plane hopping is concerned, the hopping parameters are the same whether the oxygen site above or below the Cu planes is occupied. In this way, the effect of oxygen disorder is mitigated to a large degree for $t_{\alpha \beta}$. The inter-plane hopping $t_c$, on the other hand, is strongly affected by oxygen disorder and may limit coherent transport along $c$.

We next discuss qualitatively what we may expect from the Hamiltonian given in Eqs.\ref{H0} and \ref{Hh}. First, let us set $t_c=0$ and consider decoupled honeycomb layers.  For simplicity, we first discuss the case of a single orbital on Cu and O sites. This is a canonical example of a charge transfer insulator. As a function of $D/t$ we expect a transition between an insulator with local moments for large $D/t$ to a metal for small $D/t$. Deep in the insulator phase, we expect antiferromagnetic (AF) order with spins on the Cu sites making 120 degree angle between them. In the vicinity of the transition, interesting correlation-driven physics may arise. This model has been studied in the large $D/t$ limit in the presence of doped holes away from the half-filled band limit. \cite{crepel2021new} For low hole density it is pointed out that unconventional superconductivity can arise, with a very large ratio of $T_c$ to the Fermi energy $\epsilon_F$. However, in this limit $\epsilon_F$ itself is small, so the absolute value of $T_c $ will be small. Furthermore, in the single-layer case, doping can be achieved by gating, which minimizes the effect of disorder. In contrast, in the case of LK-99, doping away from $x=1$ involves some Cu sites being replaced by Pb or some additional Cu substitution on Pb sites. This creates disorder in the active sites which is detrimental to unconventional superconductivity. Instead of doping, it may be more interesting to consider the metallic state near the metal-insulator transition to see whether some of the interesting ideas proposed in ref. \cite{crepel2021new} persist.

It is not known what the critical $D/t$ is for the honeycomb lattice. More is known about the closely related Hubbard model on a triangular lattice. Numerical studies show that the critical $U/t$ is 8.5 and interestingly a spin liquid phase exists for an intermediate range $8.5<U/t<10.6$ before the 120-degree ordered AF appears 
\cite{szasz2020chiral}. Recent DMRG calculations give evidence that the spin liquid phase is a chiral spin liquid \cite{szasz2020chiral}. In that case, one may entertain the idea of the appearance of a chiral SC on the metallic side of the transition. Experimentally, the organic salts are believed to be described by the Hubbard model on a triangular lattice at half-filling \cite{kanoda2011mott}. The metal-insulator transition can be tuned by pressure or anion substitution. It is found that superconductivity with a $T_c$ of about 10K generally appears between the insulator and the metal. It will be interesting to find out the corresponding phase diagram for the single orbital honeycomb model. It is possible that our estimate of the ratio $D/t \approx 5$ places us just on the metallic side of the metal-insulator transition where unconventional SC may be expected. We should add that in the organics, the effective hopping $t$ is estimated to be 57 meV for the anion X=Cu(CN)3 of the $\kappa$-(BEDT-TTF)$_2$X family \cite{mckenzie1998strongly}. Our estimate of the Cu-O hopping parameter $t \approx 80 meV$ is somewhat larger but of a similar order of magnitude.
10K is a far cry from room temperature, but our point is simply that the effective low-energy Hamiltonian and the parameter values we propose can place us in an interesting intermediate correlation regime where exotic physics may emerge. It is not known what the role of having two orbitals per site will play, or how different the honeycomb model is different from the triangular lattice Hubbard model.

Finally, we put back the interlayer hopping between the O orbitals, $t_c$. Given its relatively large value, it will tend to drive the system into a metallic state which will be three-dimensional. Nothing is known about this state near the metal-insulator transition or its potential to host a high-temperature SC.

In conclusion, based on the LDA band calculations, we have identified the low-energy hole excitations and we propose an effective Hamiltonian where Cu and O occupy the A and B sites of a stack of buckled honeycomb lattices. The key parameters are the in-plane hopping integral between Cu and O, the inter-layer hopping between O and the energy $D$ which separates the Cu and O onsite energies. The hopping energies and $D$ are well balanced. We propose that this system should be treated as a charge transfer system \cite{zaanen1985band} which may be close to the metal-insulator transition and has the potential of exhibiting novel physics and unconventional SC. However, the energy scale is small, considerably smaller than the energy scale in the cuprates and more comparable to the organics.  Ordinarily one would not expect very high $T_c$ with these parameters, but not much is known about systems such as these with intermediate correlations and more work is surely needed. We plan to explore some versions of these models with numerical methods.

After the completion of the work, we learn about the paper by M. Hirschmann and J. Mitscherling \cite{hirschmann2023tight} who also focused on the Cu-O bands and made tight binding fits to the LDA results. The parameter they extract for the 4-band model (model D) are similar to what we estimate with our poor man's method. %{\red In model C they assume $\Delta = 0$ and band 12 
%are equal mixing of Cu and O. In model D, they assume band 12 is purely Cu and get $\Delta = 400meV, t = 50meV$. Yang Zhang uses the actual orbital weight from his DFT to fit and find parameters in between Johannes' model C and D.} 
They focused on interesting topological aspects of these bands, whereas our emphasis is on treating this as a charge-transfer system. We also note the papers by Si et al \cite{si2023pb} and by Mao et al \cite{mao} who also provided tight binding fits to the Cu-O bands and discussed charge transfer. There are other proposed models based on the honeycomb lattice, but the B sublattice is occupied by Pb1 instead of O \cite{tavakol2023minimal}. 
We believe the Pb1 orbitals do not capture the low energy physics, at least according to LDA band calculations.

\subsection*{Acknowledgements}

 PL  acknowledges support by DOE (USA) office of Basic Sciences Grant No. DE-FG02-03ER46076. ZD acknowledges support from the Gordon and Betty Moore Foundation.

\bibliography{ref}

\end{document}